\documentclass{svproc}
\usepackage{url}
\usepackage{graphicx,bm,color,amssymb,amsmath}
\usepackage{slashed}
\usepackage{graphicx}
\usepackage{wrapfig}

\begin{document}
\mainmatter              
\title{Constraining the strength of $U(1)_A$ symmetry breaking using 2-flavour non-local NJL model}
\titlerunning{Constraining the Strength of $U(1)_A$ symmetry breaking}  
%
\author{Mahammad Sabir Ali\inst{1} \and Chowdhury Aminul Islam\inst{2}\and Rishi Sharma\inst{1}}
\authorrunning{M. S. Ali et al.} 
%
%
\institute{Department of theoretical Physics, Tata Institute of Fundamental Research, Homi Bhabha Road, Mumbai 400005, India\\
\email{sabir@theory.tifr.res.in},
\and
School of Nuclear Science and Technology, University of Chinese Academy of Sciences, Beijing 100049, China}

\maketitle              

\begin{abstract}
In presence of magnetic field we have done a systematic analysis to constrain the strength of 't Hooft determinant term using LQCD data within a non-local version of two flavour NJL model. Topological susceptibility, being related to the axial symmetry, have also been calculated and compared with LQCD results to further validate our calculation.
\end{abstract}
\section{Introduction}
With the help of the Nambu--Jona-Lasinio (NJL) model one can study the chiral properties of QCD~\cite{Nambu}. In the NJL like model the chiral condensate which breaks the chiral symmetry spontaneously, increases with magnetic field (eB) for all temperature (T), termed as magnetic catalysis (MC). Whereas lattice QCD simulation obtained decrease in chiral condensate around the crossover T as eB increases which is termed as inverse magnetic catalysis (IMC)~\cite{Bali}. Non-local extension of NJL model is very interesting as it captures some aspects of the asymptotic freedom of QCD through the non-local form factor and thus can reproduce the IMC effects automatically~\cite{Gomez,Pagura}. Our main goal is to constrain the strength of the $U(1)_A$ symmetry breaking 't Hooft determinant term, which is usually assumed to be equal to that of the axial $U(1)_A$ symmetric term both in local and non-local versions, in non-local NJL model using lattice QCD results. Along with the phase diagram in $T-eB$ plane to check our model predictability we also explored topological susceptibility ($\chi_t$) with the fitted strength of $U(1)_A$ symmetry breaking term and compare with the available LQCD data.
\section{Formalism}
The most general non-local NJL Lagrangian can be written as~\cite{Frank,Gomez}
\begin{eqnarray}
	{\cal L}_{\rm NJL}&=&\bar{\psi}\left( i\slashed{\partial}-m\right)\psi+{\cal L}_1+{\cal L}_2, \hspace{0.4cm}{\rm with}\nonumber\\
	{\cal L}_1&=&G_1\left\{j_a(x)j_a(x)+j_b(x)j_b(x)\right\}\hspace{0.4cm} {\rm and}\nonumber\\
	{\cal L}_2&=&G_2\left\{j_a(x)j_a(x)-j_b(x)j_b(x)\right\},\nonumber
\end{eqnarray}
 where $j_{a/b}(x)$ are the non-local currents, given by
\begin{center}
	$j_{a/b}(x)=\int d^4z\ {\cal G}(z)\bar{\psi}\left(x+\frac{z}{2}\right)\Gamma_{a/b}\psi(x-\frac{z}{2})$ and\\ 
	$\Gamma_a=(\mathbb{I},i\gamma_5\vec{\tau})$ and $\Gamma_b=(i\gamma_5,\vec{\tau})$.
\end{center}
${\cal G}$(z) is the non-locality form factor. As ${\cal L}_2$ explicitly breaks $U(1)_A$, the symmetry of the Lagrangian (with $m=0$) is $SU(2)_V\times SU(2)_A\times U(1)_V$. This symmetry only allows the $\langle\bar{\psi}\psi\rangle$ condensate, which depends on $(G_1+G_2)$. But in presence of isospin chemical potential ($\mu_I$) and/or $eB$, the $SU(2)$ symmetry is broken and one can have $\langle\bar{\psi}\tau_3\psi\rangle$ which has a $(G_1-G_2)$ dependence. This motivates one to parameterize $G_1$ and $G_2$ as $G_1=(1-c)G_0/2$ and $G_2=cG_0/2$ where $c=1/2$ corresponds to the standard NJL model. The parameters of our model is fitted to obtain physical quantities like pion mass, decay constant etc. in absence of $\mu_I$ or $eB$.

Due to Lorentz symmetry the Fourier transformation of $\cal G$(z), $g(p^2)$ can only depends on $p^2$. We have considered $g(p^2)$ to be Gaussian in nature
\begin{center}
	$g(p^2)=\exp[-p^2/\Lambda^2]$.
\end{center} 
In presence of $eB$ the non-local currents should transform as~\cite{Pagura},
\begin{center}
	$j_{a/b}(x)\rightarrow\int d^4z\ {\cal G}(z)\bar{\psi}\left(x+\frac{z}{2}\right)W^{\dagger}(x+z/2,x)\Gamma_aW(x,x-z/2)\psi(x-\frac{z}{2})$ \\
	\vspace{0.2cm}
	with $W(s,t)=P\ \exp\left[-iQ\int_{s}^{t}dr_\mu A_\mu(r)\right].$
\end{center}
The bosonized action, after integrating out the fermionic fields, becomes
\begin{center}
	$S_{\rm bos}=-\rm \ln \det({\cal D})+\int d^4x\left[\frac{\sigma^2(x)}{2G_0}+\frac{\Delta\sigma^2(x)}{2(1-2c)G_0}\right]$,
\end{center} 
The inverse of fermionic propagator is given by~\cite{Pagura},
\begin{eqnarray}
{\cal D}_{\rm MF}(x,x')&=&\delta^{(4)}(x-x')(-i{\slashed\partial}-QBx_1\gamma_2+m)+ {\cal G}(x-x') \nonumber\\
&&*(\sigma+\tau_3\Delta\sigma)\exp\left[\frac{i}{2}QB(x_2-x'_2)(x_1+x'_1)\right].\nonumber
\end{eqnarray}
Using Ritus eigenfunction one obtains the Fourier transform of the above to calculate the action, which then minimised with respect to meanfields to get the observables. Finite $T$ can be incorporated using Matsubara formalism; the detailed calculation can be found in Ref.~\cite{Sabir}.

The topological term $\frac{\theta g^2}{32 \pi^2}{\cal F} \tilde{\cal F}$ breaks the CP symmetry of strong interaction. As the dynamical axion is considered to be a possible solution of strong CP problem, $\theta$ can be related to the axion fields, $\theta=a/f_a$. With a chiral rotation of the quark fields by an angle $a/f_a$ one obtains
\begin{center}
	${\cal L}_2 = 2\,G_2\left\{e^{i\frac{a}{f_a}}\mathrm{det}\bar\psi(1+\gamma_5)\psi+e^{-i\frac{a}{f_a}}\mathrm{det}\bar\psi(1-\gamma_5)\psi\right\}$.
\end{center} 
With a straight forward generalization to non-local case, one can derive the free energy $\Omega(T,eB,a)$, and the topological susceptibility is given by 
\begin{center}
	$\frac{d^2\Omega(T,eB,a)}{da^2}\bigg |_{a=0}=\frac{\chi_t}{f_a^2}.$
\end{center}

\section{Results}
Three parameters of our model are fitted to obtain observables calculated in Ref.~\cite{Brandt} (referred as B13 here) ($\langle\bar{\psi}_f\psi_f\rangle^{1/3}=261(13)(1)$ MeV, $m_\pi=135\,{\rm MeV}$ and $F_\pi=90(8)(2)\,{\rm MeV}$).

\begin{figure}[!htbp]
	\includegraphics[scale=0.27]{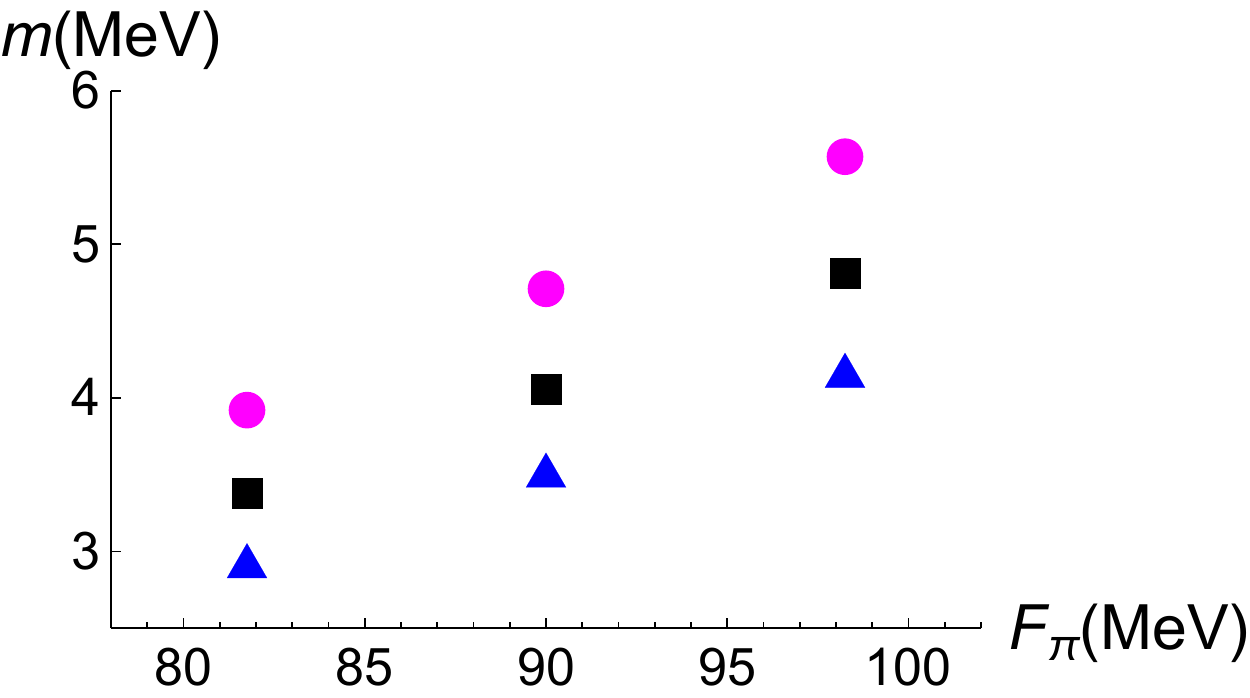}
	\includegraphics[scale=0.27]{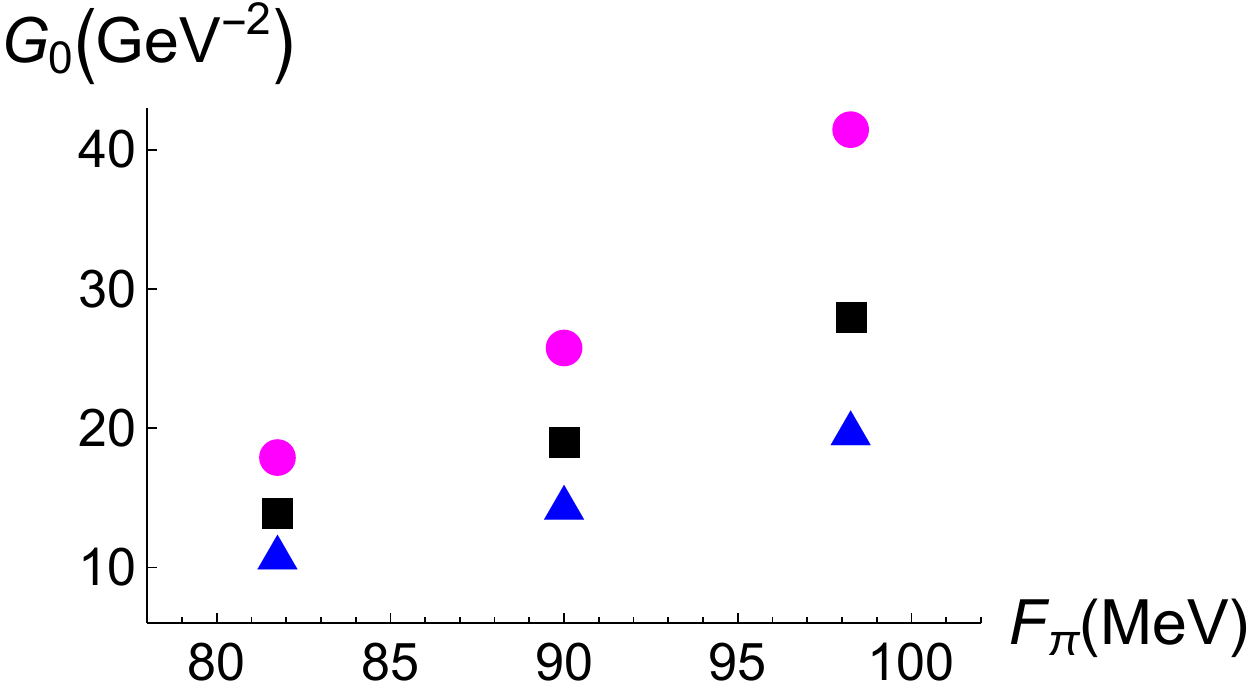}
	\includegraphics[scale=0.27]{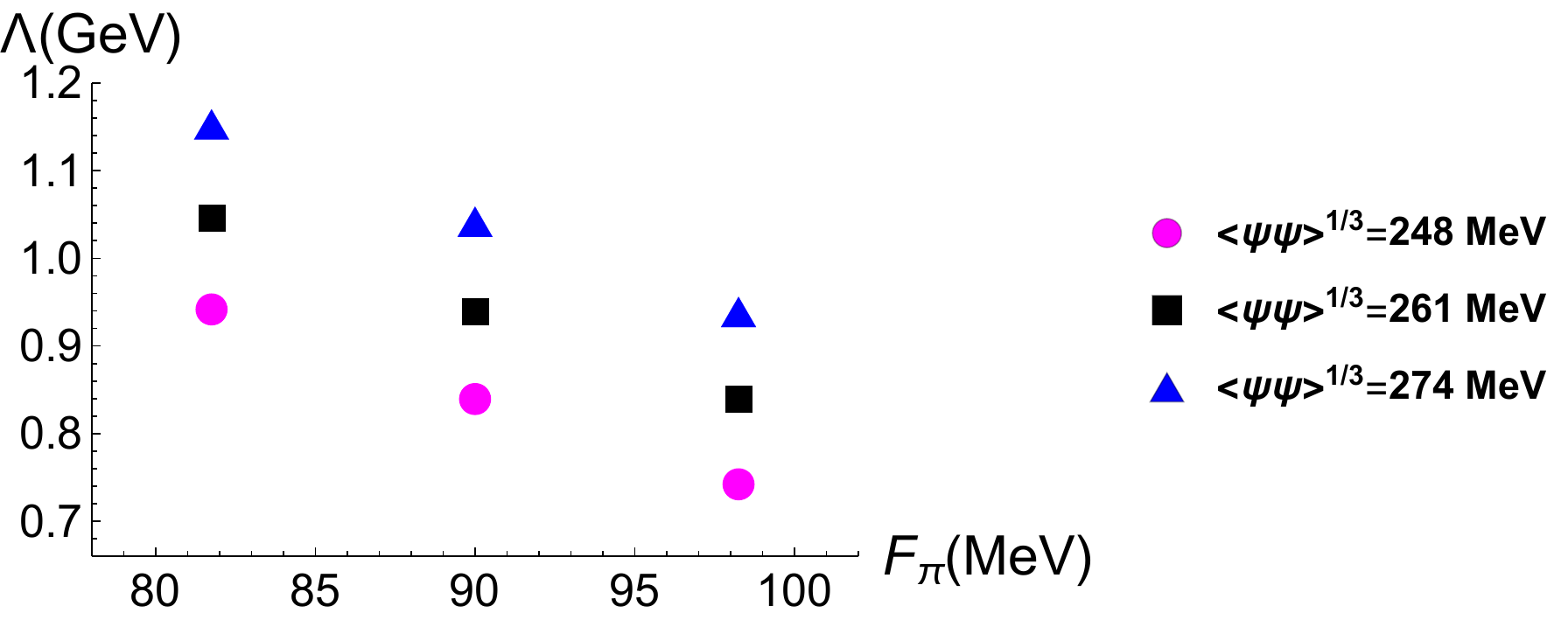}
	\caption{Range of model parameters to access the full allowed range of the
		condensate including the errors as given in LQCD~\cite{Brandt}
		(B13).}
	\label{fig:para_c_fix}
\end{figure}

Considering the error in $\langle\bar{\psi}_f\psi_f\rangle^{1/3}$ and $F_{\pi}$ we obtain a range in our model parameters. Fig.~\ref{fig:para_c_fix} represents the allowed range of our model parameters while obtaining observables of Ref.~\cite{Brandt}. Increasing $F_\pi$ results in decreasing $\Lambda$, while on the condensate the dependence is opposite.\\
\begin{figure}[!htbp]
	\includegraphics[scale=0.52]{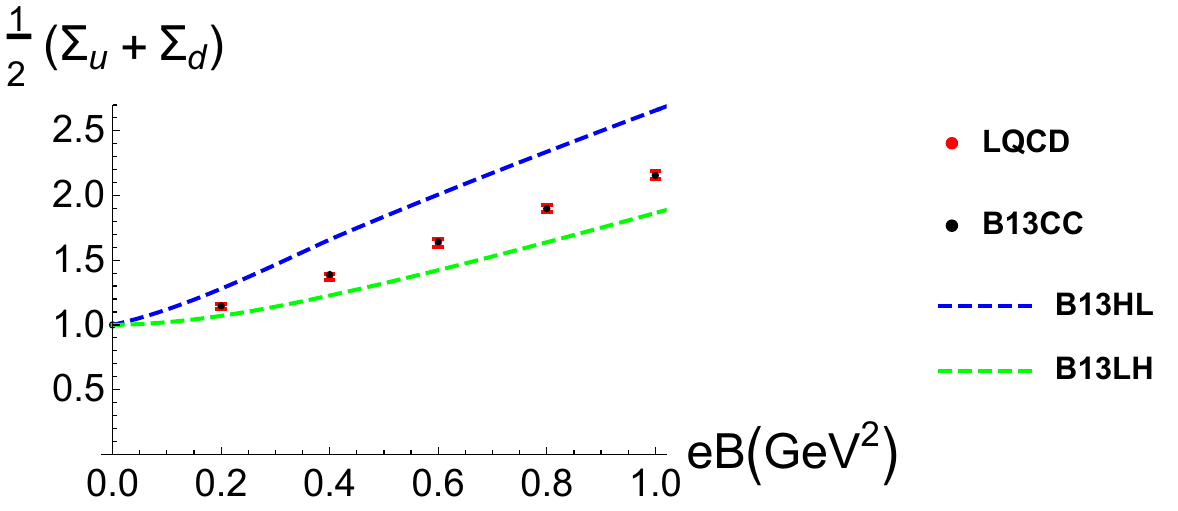}
	\includegraphics[scale=0.52]{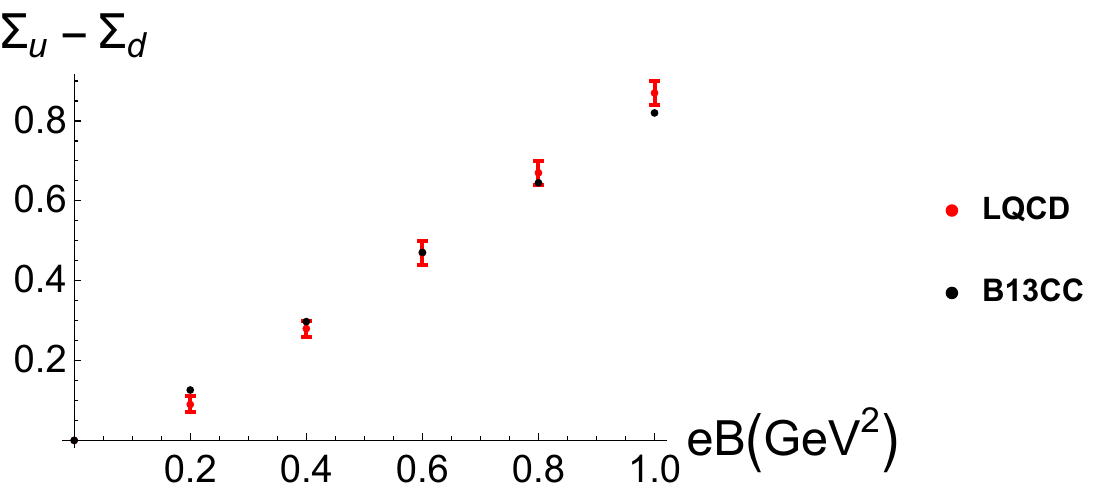}
	\caption{The condensate average (left) and difference (right) as a function of magnetic field at $T=0$ compared with LQCD~\cite{Bali} data.}
	\label{fig:FinalASB13}
\end{figure}
\begin{wrapfigure}{h}{0.4\textwidth}
\begin{center}
	\includegraphics[scale=0.46]{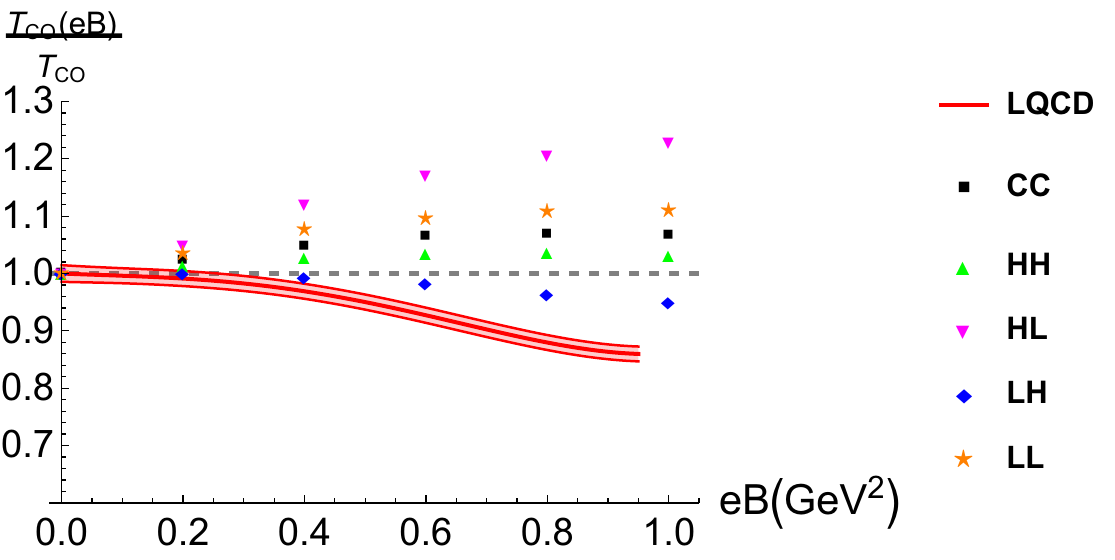}
\end{center}
\caption{The phase diagram in $T-eB$ plane for the central and all the corner parameter sets, compared with LQCD~\cite{Bali}.}
\label{fig:PDB13Com}
\end{wrapfigure}
The $T=0$ and finite $eB$ results are presented in Fig.~\ref{fig:FinalASB13} along with the comparison with lattice data~\cite{Bali}. The notations are as follows: C, H and L represents the central, high and low value, respectively and while the second last alphabet represents the  value of the condensate the last one represents that of $F_\pi$. Using $\chi^2$ fitting we fitted $c$ to obtain the best match of condensate difference (right) with the same obtained in Ref.~\cite{Bali}. The fitted $c$ value is $0.06$ with $\chi^2$/dof $=1.5$. As the $\chi^2$/dof are small we can safely assume that at $T=0$, $c$ does not depend on magnetic field. Finite $T$ behaviour of condensate average and difference can be found in Ref.~\cite{Sabir}.

Fig.~\ref{fig:PDB13Com} shows the phase diagram in $T-eB$ plane. One can clearly see that smaller value of condensate and/or larger value of $F_\pi$ produces stronger IMC effect around the crossover region. To our knowledge this is the first time that the role of $F_\pi$ on the phase diagram has been explored in effective model.

\begin{figure}[h]
	\includegraphics[scale=0.48]{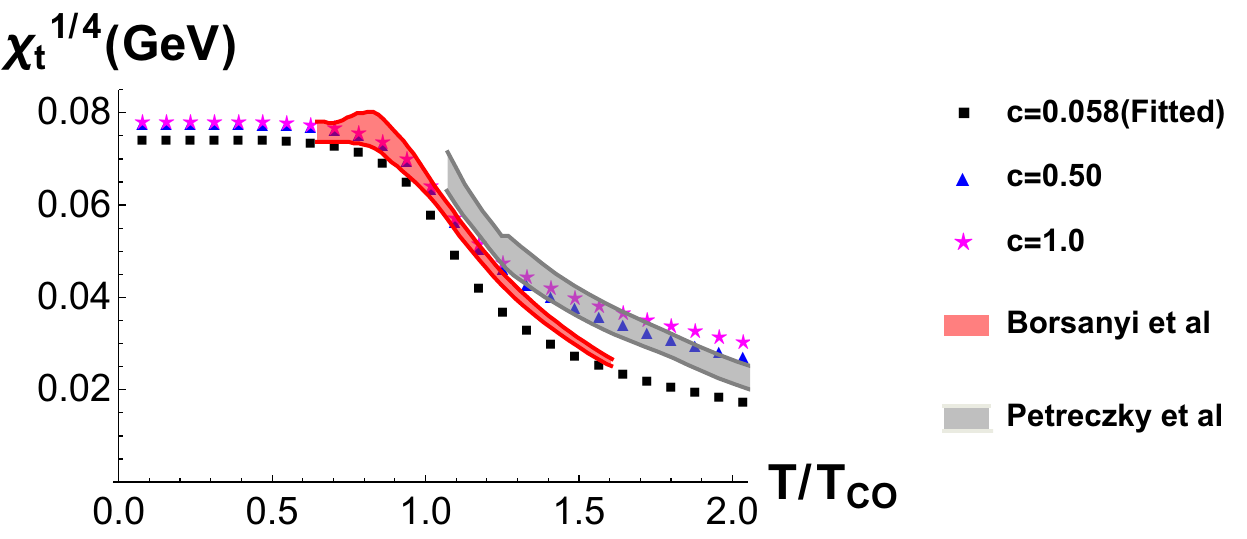}
    \includegraphics[scale=0.42]{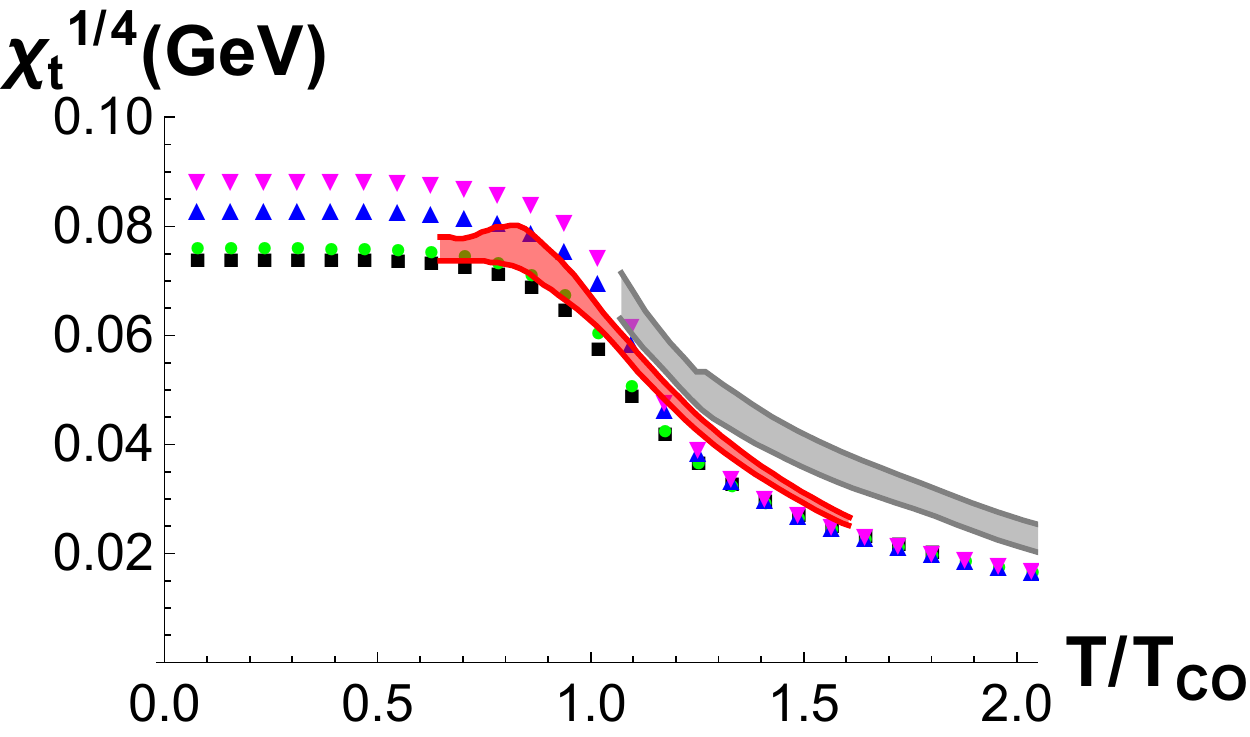}
	\caption{Topological susceptibility as a function of scaled temperature for different $c$ values (left) for the CC parameter set. eB dependence is presented on the right, dot(green), up triangle(blue) and down triangle(magenta) are for eB=0.2, 0.6, 1.0 GeV$^2$ respectively. The red and the gray bands represent lattice results from the Refs.~\cite{Borsanyi} and~\cite{Petreczky}, respectively.}
	\label{fig:top_sus}
\end{figure}

Fig.~\ref{fig:top_sus} shows our model prediction for topological susceptibility compared with two different LQCD calculation~\cite{Borsanyi,Petreczky}. Left panel shows $eB=0$ results for different $c$ values whereas the right one is for different $eB$. With $eB$ dependent topological susceptibility from LQCD (which is not available yet) one might expect $c$ to have a non-trivial $T$ and $eB$ dependence.\\
\\{\bf Acknowledgment:} CAI would like to thank TIFR and UCAS for the support.

%
%

\end{document}